\documentclass[11pt]{article}

\usepackage{graphicx}
\usepackage{amssymb}

\newcommand{\be}{\begin{equation}}
\newcommand{\ee}{\end{equation}}
\newcommand{\bea}{\begin{eqnarray}}
\newcommand{\eea}{\end{eqnarray}}
\newcommand{\nn}{\nonumber}

\newcommand{\w}{\mbox{$\omega$}}

\newcommand{\la}{\mbox{$\langle$}}
\newcommand{\ra}{\mbox{$\rangle$}}
\newcommand{\ad}{\mbox{$a^{\dag}$}}

\begin{document}

\begin{center}

\section*{A delayed choice quantum eraser explained by \newline
the transactional interpretation of quantum mechanics}

\end{center}

\vspace{0.15 in}

\begin{center}
\noindent
{\bf H. Fearn}\\

\vspace{0.1in}
\noindent
\small{Physics Department, California State University Fullerton\\
800 N. State College Blvd., Fullerton CA 92834

\noindent
{\bf email:} hfearn@fullerton.edu }\\

\vspace{0.2in}
\noindent
{\large \bf Abstract}\\

\end{center}

\noindent
This paper explains the delayed choice quantum eraser of Kim et al. \cite{shih} in terms of the transactional interpretation of quantum mechanics by John Cramer \cite{cramer, cramerbk}. It is kept deliberately mathematically simple to help explain the transactional technique. The emphasis is on a clear understanding of how the instantaneous ``collapse" of the wave function due to a measurement at a specific time and place may be reinterpreted as a relativistically well-defined collapse over the entire path of the photon and over the entire transit time from slit to detector. This is made possible by the use of a retarded offer wave, which is thought to travel from the slits (or rather the small region within the parametric crystal where down-conversion takes place) to the detector and an advanced counter wave traveling backward in time from the detector to the slits. The point here is to make clear how simple the transactional picture is and how much more intuitive the collapse of the wave function becomes if viewed in this way. Also, any confusion about possible retro-causal signaling is put to rest. A delayed choice quantum eraser does not require any sort of backward in time communication. This paper makes the point that it is preferable to use the Transactional Interpretation (TI) over the usual Copenhagen Interpretation (CI) for a more intuitive understanding of the quantum eraser delayed choice experiment. Both methods give exactly the same end results and can be used interchangeably.\\

\noindent
{\bf PACS codes:}  03.65.Ta, 03.65.Ud, 42.50.Dv \\

\noindent
{\bf Key Words:} Transactional interpretation, advanced waves, delayed choice, quantum eraser \\

\noindent
\subsubsection*{Complementarity, which path information and quantum erasers}

Feynman 1965, in his famous lectures on physics \cite{fey} stated that the Young's double slit experiment {\em contains the only mystery} of quantum mechanics. We may see interference, or we may know through which slit the photon passes, but we can never know both at the same time. This is what is commonly referred to as the principle of complementarity. We say two observables are {\em complementary} if precise knowledge of one implies that all possible outcomes of measuring the other are equally likely. The fundamental enforcement of complementarity arises from correlations between the detector and the interfering particle in a way that show up in the wave function for the system. It is not, as some undergraduate text books would have you believe, a consequence of the uncertainty principle. The Heisenberg uncertainty relation is a consequence of complementarity, not the other way around. There have been many {\it gedanken} (German for thought) experiments over the years to show complementarity.  The most famous are the Einstein recoiling slit, Feynman's light scattering scheme both discussed in Feynman's lectures on physics \cite{fey} and Wheeler's delayed choice experiment \cite{wheel}.

\begin{figure}
  \begin{center}
   \includegraphics[height=1.75in, width=4.5in]{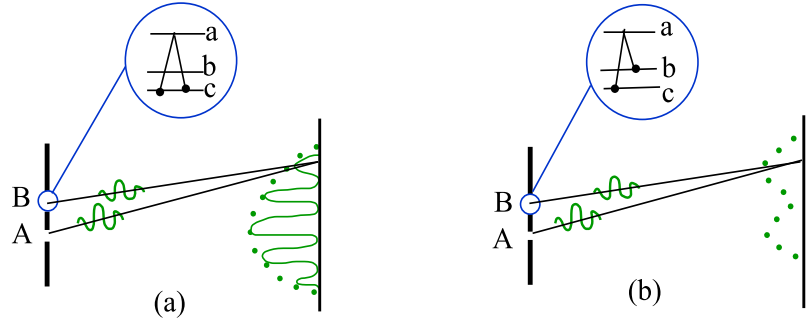}\\
   \caption{The figure shows the Scully Druhl quantum eraser 2 slit arrangement. Two 3-level atoms are in place of the two slits. A laser excites either atom to the upper level $a$ which may then decay to level $b$ or $c$. If the atom decays to level $c$, the ground state, then there will be interference since there is no way to distinguish between the two atoms and so no which path information. See figure (a).  The green dots represent the single slit diffraction pattern. The solid line is the intensity detected. If the atom decays to level $b$, then there is which path information and there will be no interference pattern as in figure (b). The drawings are simplified.  }
  \end{center}
\end{figure}

\noindent
Of particular interest here is the delayed choice quantum eraser {\it gedanken} experiment by Scully and Druhl  1982 \cite{druhl}. This work described a basic quantum eraser experiment and a delayed choice quantum eraser arrangement. The basic quantum eraser experiment is described using two 3-level $\Lambda$--type atoms \cite{scullybk}, in the place of two slits.  See Fig. 1. The atoms start off in the ground state and then a laser pulse comes in and excites either atom $A$ or $B$. The excited atom then decays and emits a signal photon. Interference fringes are sought between these signal photons on a screen some distance away. Let the identical 3-level atoms have one upper level $a$ and two lower levels $b$ and $c$. The laser excites one of the atoms up to the level $a$ but the atom can de-excite to either state $b$ or $c$. If both atoms start off in the ground state $c$, there are two possibilities. The excited atom decays and falls back to level $c$, so the excited atom becomes indistinguishable from the other atom which was not excited. In this case we would expect to see an interference pattern since there is no which path information. In the second case, the excited atom drops to level $b$ which is distinguishable from level $c$. In this case we have which path information and we would get no interference pattern. That describes the basic quantum eraser.\\

\noindent
For a delayed choice quantum eraser \cite{druhl}, the 3-level atoms change to 4-level atoms with levels $a,b,c,d $, with $d$ the ground state. See Fig. 2. Instead of one exciting laser pulse there are two closely spaced pulses, which will both go to the same atom. The first laser pulse excites either atom $A$ or $B$ from the ground state $d$ to the upper level $a$. The excited atom then spontaneously decays to $c$ emitting the signal photon. The second laser pulse then excites the atom from level $c$ to level $b$, which then decays with the emission of a lower energy {\em idler} photon to the ground state. Now the atoms are inside a cleverly constructed cavity with a trap door separating them. The cavity is transparent to the signal photons and laser light but strongly reflects the idler photons. There is a detector capable of detecting the {\em idler} photons only near atom $A$. The trap door will prevent the {\em idler} photon from $B$ being detected. Now we have a choice whether to open the trap door or leave it closed.  The signal photon detection is now correlated with the idler photon detection. The experiment has become a delayed choice quantum eraser, whether we see interference or not will depend on whether we leave the trap door open or closed.  If the trap door is closed and we detect an idler photon, we know that atom $A$ was excited. If we do not detect a photon then atom $B$ was excited, either way we have which path information that will destroy the fringes. If the trap door is open, then we no longer have which path information since either atom could have emitted the idler photon. In principle the decision, to leave the trap door open or closed, can be made {\em after} the signal photons have been detected. The {\em paradox} is, how does the signal photon know which pattern to make, a single slit diffraction pattern or a two-slit interference pattern, if we have not yet decided to leave the trap door open or closed?\\

\begin{figure}
  \begin{center}
   \includegraphics[height=2.0in, width=3.0in]{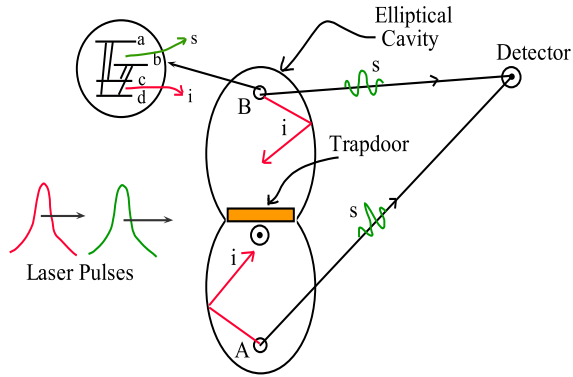}\\
   \caption{The figure shows the Scully Druhl delayed choice quantum eraser. Two 4--level atoms, labeled $A$ and $B$ are
   in place of the two slits.
   The atoms are inside a double elliptic cavity with a shared focus. Both atoms and the idler detector are located at foci. The first laser pulse excites either atom to the upper level $a$, which then decays to level $c$, emitting a signal photon (green), which leaves the cavity. The second laser pulse immediately excites the same atom from level $c$ to level $b$, which then decays to the ground state $d$, emitting an idler photon (red). The idler photons cannot leave the cavity. When closed, the trapdoor prevents idler photons from atom $B$ being detected. When the trapdoor is open, the cavity detector may detect idler photons from either atom. }
  \end{center}
\end{figure}

\noindent
Englert, Scully and Walther \cite{nature} in 1991 constructed a very nice atom interference {\it gedanken} experiment that shows the physics in a very straightforward manner, although the experiment would be extremely difficult to perform in practice. Soon afterward in 1993, a polarization experiment by Wineland's group \cite{itano}, was the first to demonstrate an actual realization of the Scully--Druhl quantum eraser {\it gedanken} experiment. They used mercury ions in trap as the two ``atoms" and observed linear $\pi$ and circularly $\sigma$ polarized light. Choosing to detect linear polarized light, corresponded to the case that the ions in a trap were in the same initial and final state. This implies that there was no which path information and so there was interference. Choosing to observe circular polarized light, corresponded to the situation that the ions were in distinguishable end states after scattering a photon, so which path information was available and hence there was no interference.  You could choose to observe interference or not depending on whether you chose to observe linear or circularly polarized light. \\

\noindent
There have been many quantum optics experiments involving two photon entangled states and quantum eraser arrangements to demonstrate the complementarity arguments above. Three of the better ones are \cite{holladay,herzog,chiao}.  One experiment in particular by Zeilinger's group \cite{song} is worthy of a special note. The arm lengths in their apparatus were very long, between 55 m up to 144 Km. They point out that there is no possible communication between one photon and the other in the entangled pair because of the space-like separation between them and they assume no faster-than-light communication is possible. \\

\noindent
The most famous real experiment of the delayed choice type is that by Kim et al. \cite{shih}, using parametric down conversion entangled photons. It has drawn considerably more press than any other experiment of this type and even has a couple of online animations \cite{utube}. We choose to present our case for the transactional interpretation of quantum mechanics using the Kim experiment as our example, but any of the delayed choice quantum erasers would work just as well.

\subsubsection*{Introduction to the Transactional Interpretation of Quantum Mechanics}

\noindent
 The transactional interpretation of quantum mechanics was proposed by John Cramer \cite{cramer} in a review article in 1986 and a short overview in 1988 \cite{cramer1}. More recently Cramer has written a book \cite{cramerbk} which should become available early in 2016. It is a way to view quantum mechanics that is very intuitive and easily accounts for all the well known quantum paradoxes, Einstein Rosen Podolsky (EPR ) experiment \cite{cramer4}, which-way detection and quantum eraser experiments, \cite{cramer3,cramer5}.
Unfortunately, it has garnered little support over the years and has fallen off the radar. It deserves
 a much broader dissemination and part of the motivation to publish this paper was to bring Cramer's ideas, and the advanced wave concept, to the attention of the younger generation of physicists, who may not have heard of them before. The advanced wave is a standard solution of relativistic wave equation and was utilized by such notable physicists as Dirac, Wheeler, Feynman, Davies, Hoyle and his doctoral student Narlikar.  The direct particle interaction theory (which uses advanced waves, traveling backward in time) was used by Wheeler, Feynman, Schwinger, Hoyle and Narlikar. The direct particle interaction does away with the idea of a field, the vacuum field then would be truly empty, with zero energy, as Feynman believed. Frank Wilczek recounts a conversation with Feynman \cite{wilz}. \\
 
 \begin{quote}
Around 1982, I had a memorable conversation with Feynman at Santa Barbara.  Usually, at least with people he didn't know well, Feynman was ``on" -- in performance mode.  
 But after a day of bravura performances he was a little tired and eased up.  Alone for a couple of hours, before dinner, we had a wide-ranging discussion about physics.  Our conversation inevitably drifted to the most mysterious aspect of our model of the world-- both in 1982 and today-- the subject of the cosmological constant.  
 (The cosmological constant is, essentially, the energy density of empty space.  Anticipating a little, let me just mention that a big puzzle in modern physics is why empty space 
 weighs so little even though there's so much to it.)
	I asked Feynman, ``Doesn't it bother you that gravity seems to ignore all we have learned about the complications of the vacuum?" To which he immediately responded,
  ``I once thought I'd solved that one."
	Then Feynman became wistful.  Ordinarily he would look you right in the eye, and speak slowly but beautifully, in a smooth flow of fully formed sentences or even paragraphs.  Now, however, he gazed off into space; he seemed transported for a moment, and said nothing.
	Gathering himself again, Feynman explained that he had been disappointed with the outcome of his work on quantum electrodynamics.  It was a startling thing for him to say, because that brilliant work was what brought Feynman graphs to the world, as well as many of the methods we still use to do difficult calculations in quantum field theory.  It was also the work for which he won the Nobel Prize.
	Feynman told me that when he realized that his theory of photons and electrons is mathematically equivalent to the usual theory, it crushed his deepest hopes.  He had hoped that by formulating his theory directly in terms of paths of particles in space--time -- Feynman graphs -- he would avoid the field concept and construct something essentially new.  For a while he thought he had.
	Why did he want to get rid of fields?  ``I had a slogan,"  he said.  Ratcheting up the volume and his Brooklyn accent, he intoned it:

	The vacuum doesn't weigh anything [dramatic pause] because there's nothing there!
\end{quote} 
 
 \noindent
Experimental observations show that the vacuum energy density is in fact very close to zero. 
To calculate the vacuum energy in quantum field theory, we must admit that spacetime is probably not a continuum but rather has a  discrete nature, at quantum dimensions, and only sum the zero-point energies for vibrational modes having wavelengths larger than, the Planck length ( $10^{-35}$ m) and less than or equal to the size of the universe (diameter approx. $8.8 \times 10^{26}$m). This gives a ridiculously large but finite vacuum energy density of about $10^{111}$ J$m^{-3}$ or in terms of mass density $ 10^{94}$ Kg$m^{-3}$. 
Clearly, no where near the experimentally observed value for energy density near zero and mass density, near the critical value of $10^{-26}$ Kg$m^{-3}$. The quantum field theory vacuum mass density is about 120 orders of magnitude too large -- rather embarrassing really.\\

\noindent
{\bf Absorber theory and Advanced Waves and Direct Particle Interaction Theory}\\

\noindent
The idea of advanced waves in classical electrodynamics started with Dirac \cite{dirac} in 1938 and his derivation of the radiation reaction of a charged accelerated particle. Advanced waves travel backward in time and are a perfect way to allow for action-at-a-distance.  A remote particle can interact with a local source particle by absorbing retarded waves from the source in the future and in response, emits an advanced wave which travels backward in time and interacts with the source immediately, at the instant the retarded wave was emitted. This is a direct particle interaction and does not require the presence of a field. This direct particle interaction conserves momentum. Dirac assumed an advanced wave, in his radiation reaction calculations, but gave no physical explanation as to where it came from. Later Wheeler and Feynman \cite{WF} wrote papers in 1945 and 1949 on absorber theory, which was their attempt to give a physical description of the origins of the advanced waves introduced by Dirac. An added motivation was to try and remove the self energy from the electron, but that was not entirely successful, as Wilzcek recounts above. The radiation reaction could be accounted for without self interaction, but at the quantum level self-interaction became unavoidable for charge renormalization, electron --positron pairs are still required to shield the infinite negative bare mass. With an upper bound (Rindler horizon caused by an accelerating expansion ) and lower length cutoff (Schwarzschild radius of a particle about $10^{-45}$cm) , the standard renormalization procedure can be applied to the direct particle interaction approach, which is then no less suitable than conventional field theory, but has no cosmological constant problem. There are no classical divergences if the self-interaction is non-quantized.
Feynman's PhD thesis included the path integral approach to non-relativistic quantum mechanics, which was used to describe how to quantize the {\em direct particle interaction } of absorber theory \cite{space48}. Paul Davies later generalized these classical results for the relativistic case of absorber theory \cite{davies1, davies2}. Hoyle and Narlikar also worked on the relativistic absorber theory \cite{HN0}. There are now three different models for absorbers which have slightly differing advanced wave behavior. Wheeler-Feynman \cite{WF}, Csonka \cite{csonka}  and Cramer \cite{cramer}. These models differ with regard to what exactly happens when there is a less than perfect absorber present. They are discussed in the very readable paperback by Nick Herbert \cite{nick}. So far, we have a working theory for classical electrodynamics and now for QED. Hoyle and Narlikar have also generalized Einstein's theory of gravitation by using a direct particle interaction. Their theory reduces to Einstein's general relativity in the limit of a smooth fluid approximation, in the rest frame of the fluid. This has the benefit of completely incorporating Mach's principle as a radiative interaction between masses, \cite{HNbook}. Cramer spells out the general quantum version of the theory applicable to all systems not just electrons \cite{cramer2,cramer6}.\\

\begin{figure}
  \begin{center}
   \includegraphics[height=2.0in, width=2.0in]{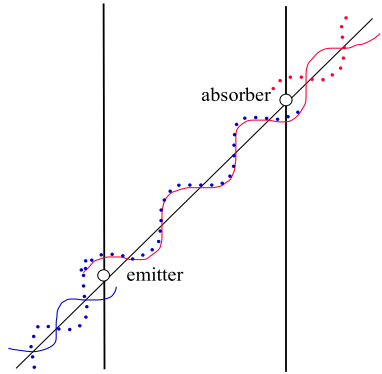}\\
   \caption{Cramer's {\em wiggle} diagram. The figure shows a plane-wave transaction between an emitter and an absorber particle. The black vertical lines are the world-lines for each particle. Waves from the emitter are solid lines, waves from the absorber are dotted. The retarded waves are red for both emitter and absorber and the advanced waves are blue.  Red retarded waves move up toward the right. Blue advanced waves move downward to the left. Note that along the path between the emitter and absorber the waves add constructively  but before the emitter and after the absorber the waves destructively interfere.
  }
  \end{center}
\end{figure}

\noindent
{\bf Transactional Interpretation}\\

\noindent
For an interaction to take place between two particles, emitter and the absorber, Cramer says the emitter must send out an offer wave. This offer wave would be half an advanced and half a retarded wave going out in all directions looking for an absorber, something to interact with.  When the retarded offer wave reaches the absorber, that particle sends out a counter wave, also half retarded and half advanced. See Cramer's wiggle diagram, Fig. 3. The advanced counter wave would travel backward in time, along the exact incident path of the original retarded offer wave (it is the complex conjugate of the retarded offer wave), thus constructive interference takes place along the path between the particles. In the one spatial dimension drawn in Fig. 3, the advanced counter wave reaches the emitter particle at the exact time when the retarded offer wave was emitted. This enables the advanced wave from the absorber to exactly cancel with the advanced wave from the emitter at the location of the emitter. Likewise, the retarded wave from the emitter will cancel the retarded wave from the absorber at the location of the absorber. Only the retarded wave from the emitter and the advanced wave from the absorber along the adjoining path are enhanced by the superposition, they do not cancel out. These waves represent the interaction between the particles.\\

\noindent
In three spatial dimensions things are a little more complicated. Advanced and retarded waves travel in all directions not just in the direction of one absorber. Retarded waves carry on into the future and maybe absorbed at some later point in time. An advanced wave travels backward in time to the {\em big bang}. At this point it is reflected and will move forward in time {\em as an advanced wave} identical to, and $\pi$ out of phase with, the incident advanced wave. This will produce a cancellation at every point along the world-line back to the point of emission of the wave. All advanced waves therefore cancel out, \cite{arrow}.
Note that the waves are assumed to travel at speed $c$ the speed-of-light in a vacuum, although the advanced wave is traveling
backward in time, or with -$t$ \cite{cramer2}.
Basically, in quantum terms, the regular wave function is the offer wave, (or at least the retarded wave part that does not cancel out) the complex conjugate wave function is the confirmation wave (the advanced part moving between the absorber going back in time to the emitter) and together they give a {\em handshake} \cite{cramer6}, which allows an interaction to take place.\\

 \noindent
 Recently Kastner \cite{ruth} has expounded the virtues of the transactional method with an additional twist allowing for {\em free will}. There are many examples of the use of the transactional method in the book and it is well worth a read. In this paper we make no distinction between the original Cramer Transactional Interpretation (TI) and the Kastner version of Possibilist Transaction Interpretation (PTI). Kastner's approach \cite{kastner},
 \begin{quote}
  ``is to consider a growing emergent universe in which the future is not set in stone but is actualized from an underlying substratum of quantum possibilities.''
  \end{quote}

\noindent
Cramer's approach means (from the authors view point) that the future is set, the past, present and future may all coexist and we simply have the illusion of flowing through time. To avoid confusion, we quote Cramer on his own interpretation \cite{email};

\begin{quote}
``Let me give an example.  When you use your cash card at the grocery store to pay for your purchases, the electronic handshake that occurs between the bank and the cash register insures that money is ``conserved'' and is neither created nor destroyed, but it does not determine what you elected to purchase.  The same is true with quantum transactions, which guarantee the conservation laws but do not determine the future.  The real difference between Kastner's PTI and my TI is that for her, offer and confirmation waves exist as objects only in some multidimensional Hilbert space.  In the TI the waves exist in real 3+1 dimensional space.  Hilbert space was invented by theorists prone to abstraction because it was the only way they could imagine that quantum waves could be entangled.  The TI explains how they can be entangled, because the multi-particle transactions allow only those subset of the waves that satisfy the conservation laws to become real transactions.''
\end{quote}

\noindent
Others have considered a Many-Worlds Interpretation, with every possible event happening along parallel realities in order to maintain {\em free will}. Neither Kastner nor Cramer agree with the many-worlds view \cite{cramer4}.
Here, the reader is asked to make up their own mind.  This paper is concerned only with; {\em Does the transactional interpretation fit the data or not?} It is found that all the usual quantum results hold and the TI is simply an alternative point of view from the Copenhagen interpretation, and the instantaneously collapsing wave function, way of thinking.\\

\subsubsection*{The delayed choice quantum eraser by Kim et al.}
\noindent
First we briefly explain the experiment and the observed results. The experimental  arrangement can be seen in Fig. 4. An argon  laser ( $\lambda_p = 351.1$ nm) is passed through a double slit and illuminates a type II phase matching nonlinear crystal of $\beta$- Barium Borate BBO ($\beta-BaB_2O_4$)  The slit $A$ allows region $A$ of the crystal to be illuminated and slit $B$ allows only region $B$ of the crystal to be illuminated. This small region is about 0.3 mm long which we take to be the slit width $a$. The separation $d$ of the two regions is about 0.7 mm as specified in the paper \cite{shih}. So we may discuss regions $A$ and $B$ of the crystal just as well as the original 2 slits. Parametric down conversion will occur at both sites and from the one pump photon will emerge two photons, a signal and an idler.  Note that all possible frequencies are created $\nu_p = \nu_s + \nu_i$. We are selecting two of the same frequency, or equivalently, twice the pump wavelength $\lambda_s = 702.2$ nm.
The signal and idler photons represent the e-ray and o-ray of the nonlinear crystal.\\

\noindent
These photons are momentum entangled and are created essentially at the same time. The probability for a downconversion event is slight, so we may assume that there is only one entangled pair of photons in the system at any given time. Different wavelengths of signal and idler photons exit the crystal at different angles. The required wavelengths are selected by restricting the exit angle. Usually a small
range of wavelengths would be selected. For convenience we track only one wavelength, but we should bear in mind that there will be a small bandwidth of wavelengths which will affect the interference pattern of the signal photons and change the visibility of the fringes accordingly. The bandwidth can also be changed using filters in front of the detectors. The detectors will have a less than perfect efficiency which will also affect the fringe visibility. The efficiency of the detectors was not mentioned in the experiment however, and neither was the effective bandwidth.\\

\noindent
The signal photons are sent though a lens, of focal length $f$, (not specified in the paper \cite{shih}) and then focussed onto a screen where they can be detected by detector $D_0$.
The detector scans, via stepper motor, along the $x$-axis to build up a pattern. The lens is used to create the far field condition at the detector so we expect a Fraunhofer type pattern which is built up over time. The idler photons, from region $A$ and $B$ of the crystal, are sent in the direction of a Glen-Thompson prism (a wedge mirror is used in figure 4. instead) which separates them into different paths.
The idler photons from region $A$ hit BSA and are either reflected or transmitted. The reflected photons will be detected by $D_3$. The transmitted photons will be reflected by mirror MA and then either transmitted through the beamsplitter BS to detector $D_2$ or reflected by BS into detector $D_1$.
The idler photons from  $B$ hit BSB and are either reflected or transmitted. The reflected photons will be detected by $D_4$. The transmitted photons will be reflected by mirror MB and then either transmitted through the beamsplitter BS to detector $D_1$ or reflected by BS into detector $D_2$.\\

\begin{figure}
  \begin{center}
   \includegraphics[height=2.5in, width=4.0in]{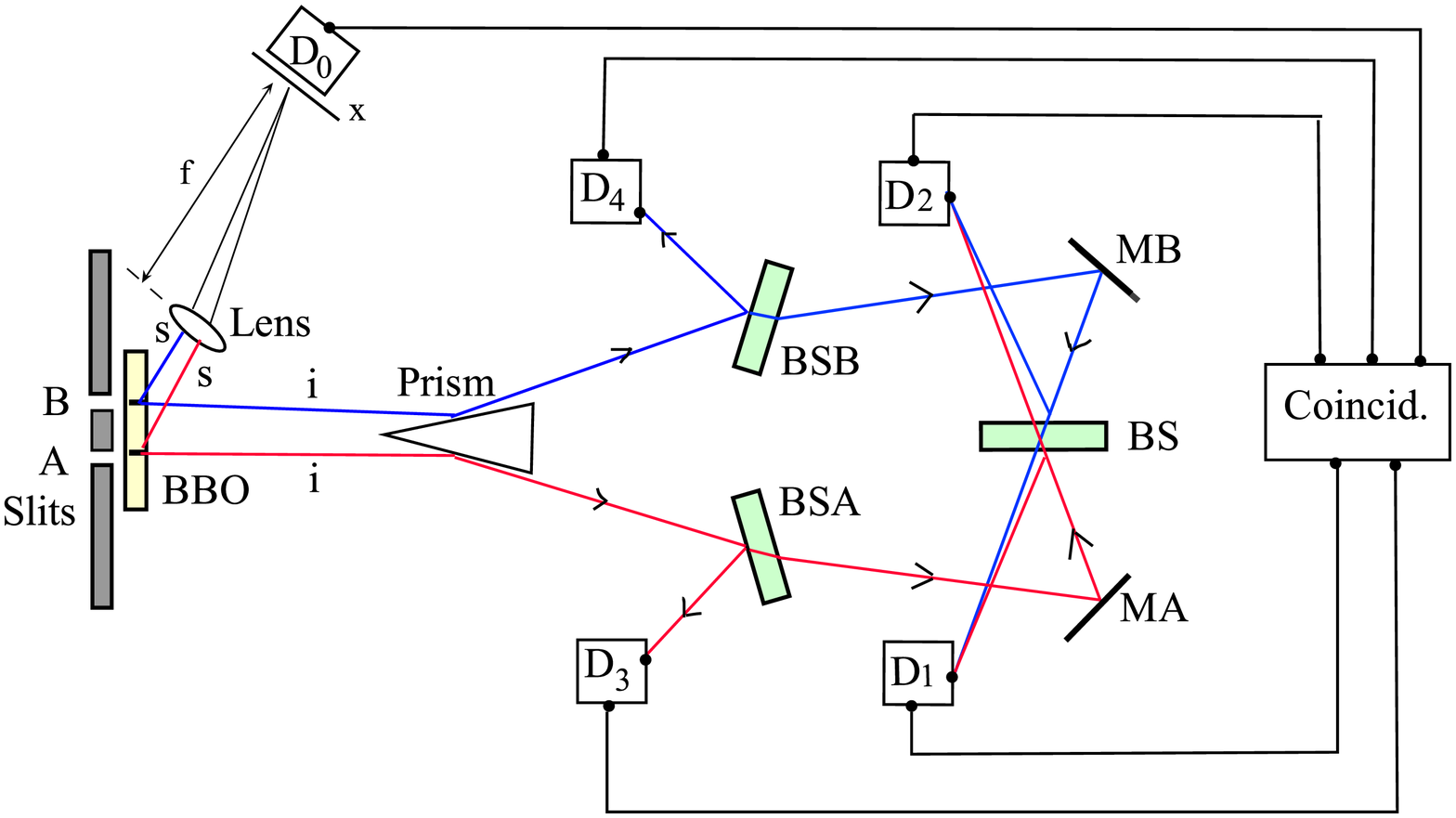}\\
   \caption{The figure shows the set up for the Kim et al. delayed choice experiment. All three beamsplitters, BSA, BSB and BS, are 50:50 lossless beamsplitters. A pump laser is incident on two slits $A$ and $B$ which also corresponds to two different small regions within a BBO ($\beta-BaB_2O_4$) crystal for parametric down conversion. We assume that the signal and idler photons are the same frequency and are both half the pump frequency.
 Photons from region $A$ are colored red and photons from region $B$ are colored blue for tracking convenience only. The signal photons (marked s) from both regions go to detector $D_0$ where an interference pattern may or may not be observed. The idler photons (marked i) from both regions are separated by a wedge mirror (a prism was used in the actual experiment) and then go to beamsplitters BSA and BSB respectively. Idler photons from $A$ alone are recorded by detector $D_3$ by reflection from BSA and idler photons from $B$ alone are detected by $D_4$ by reflection from BSB. If the idler photons are transmitted through BSA or BSB then they are mixed by the third beamsplitter BS and can be detected by either detector $D_1$ or $D_2$. The idler photons at these detectors no longer carry any which path information. All detectors then go to a coincidence counter.
 The diagram is meant to illustrate the same arm lengths for the red and blue idler photons, the reflections from the mirrors and beamsplitters {\em are not} accurately drawn with correct angles and refraction {\em is not} included.}
  \end{center}
\end{figure}

\noindent
The time of flight from the crystal to the detector $D_0$ for the signal photons is 8 ns shorter than for the idler photons which go in the direction of the beamsplitters and were eventually detected by detectors $D_3, D_4$ or by $ D_1 $ or $D_2$. The equivalent path length is approximately 2.5 m. We assume that all the detector path lengths, $D_1$ -- $D_4$, are the same and equal to 2.5 m. This path length will introduce a constant phase shift into each joint detection. It is also assumed that all mirror reflection angles are the same in both paths so that no additional phase shift differences need to be considered. Since all the phase shifts are considered equal they will cancel out and will not effect the overall interference pattern.\\

\noindent
 All the detectors are linked to a coincidence counter and the interference patterns are recorded.
The intensity pattern recorded at $D_0$ shows no interference when there is a coincidence between $D_0$ and $D_3$ or $D_4$. In these cases, we have which path information, since $D_3$ only records idler photons from slit $A$ and $D_4$ only records idler photons from slit $B$. Since the signal and idler photons come from the same region of the crystal, we would then know through which path the signal photons came and we expect no interference.\\

\noindent
When the coincidence counts are between $D_0$ and $D_1$ there is an interference pattern. The beamsplitter BS mixes the idler photons from both regions and we have now {\em erased} the which path information. There is also an interference pattern  when there is a coincidence between $D_0$ and $D_2$ but this pattern differs from the previous one by a phase shift of $\pi$. In other words if one pattern shows a co-sinusoidal interference the other will be sinusoidal. The experiment is considered a delayed choice quantum eraser since the
signal photons path length is shorter than the idler photons. It would seem that the signal photons are detected first, then we make a selection of which coincidence detections to look at, and depending on that choice we see or do not see interference of the signal photons. The paradox being, how can you influence the signal photon, basically tell it to interfere or not, by making a choice of detector $D_1 - D_4$,  \\
{\em 8 ns after} the signal photon has already been detected by $D_0$. This however is the wrong way to think about this problem. If looked at in the correct way there is no paradox.\\

\noindent
These observations can easily be explained in terms of the transactional interpretation of quantum mechanics as follows. A brief account of this experiment is given in the book by Kastner \cite{ruth}, we give a bit more detail here.

\subsubsection*{Transactional interpretation derivation}

\noindent
Let us start with a few preliminaries. The three beamsplitters in the experiment are all 50:50 lossless beamsplitters.
When a photon wavepacket goes through one of these beamsplitters there is no loss so one would expect the probability amplitude of the wave function to remain unaltered.
\bea
|\psi |^2  &=& | r \psi + t \psi |^2 \nn \\
&=& [ |r|^2 +|t|^2 + ( r^* t + r t^* ) ] |\psi |^2
\eea
\noindent
This means that the amplitude reflection and transmission coefficients obey,
\bea
|r|^2 + |t|^2 &=& 1 \nn \\
|r|^2 = |t|^2 &=& \frac{1}{2} \nn \\
r^* t + r t^* &=& 0 \nn \\
\mbox{ hence    } \;\; r &=& \frac{i}{\sqrt{2}}  \;\;\; \mbox{  and  } \;\; t = \frac{ 1}{\sqrt{2}} \; .  \label{amps}
\eea
We take all the beamsplitters to be identical for convenience. It will be assumed that each optical path length for the idler photons, is the same and any phase changes due to mirror reflections have been compensated for.
\noindent
An offer wave will go out from the slits and get absorbed by a detector. The detector will then send back an advanced wave (backwards in time) along the same path as the incident wave to the slits to {\em handshake} and confirm the interaction. Only then does the photon actually leave the slit region. The offer wave is a momentum entangled two-photon state (or bi-photon). The possible transactions will depend on the detector configuration which generates the counter wave. We will go through the process step by step. \\

\noindent
The original offer wave from the slits comes from the pump laser beam, we will take this to be,
\be
\psi = \frac{\alpha}{\sqrt{2}} \left( | A_p \ra + | B_p  \ra \right)
\ee
where the subscript $p$ stands for pump. The $\alpha$ is the single slit diffraction pattern, a sinc function of the usual kind. $A$ and $B$ stand for the photon wave functions from the two slits, of plane wave type.  Parametric downconversion inside the $\beta$-barium borate (BBO) crystal duplicates each pump photon into a signal and an idler photon.
For type I parametric down conversion, the signal and idler have the same polarization, for type II the signal and idler polarizations are perpendicular. This is of no importance here since the signal photons from regions $A$ and $B$ interfere at detector $D_0$ and both idler photons interfere at one of the four detectors $D_1$--$D_4$.
The offer wave from the 2 slits and crystal then becomes,
\be
\psi = \frac{\alpha}{\sqrt{2}} \left( | A_s \ra | A_i \ra + | B_s \ra  | B_i \ra    \right) \;\; .
\ee
\noindent
 We select both the signal and idler photons of half the pump frequency, by restricting the exit angle from the crystal. Even so there will be a small spread in frequency, and thus wavelength, which will cause the fringe visibility to be less than perfect. However, we will continue thinking of the photons wave functions as simple monochromatic plane waves  for simplicity. It is easy to generalize the end result for more than one wavelength.

 The time dependent, correlation function calculation can be found in Appendix A. This is for comparison with the TI approach taken below. We skip the details of the parametric downconversion process in what follows, but they can be found in \cite{ref1,ref2,ref3, scullybk} and these results are used in the Appendix A calculation. The first reference refers to 5 basic quantum experiments and has simple theory accessible to undergraduates \cite{ref1}. The second reference has more theory but still some experiment, and is geared more for graduates and researchers \cite{ref2} and the last two reference is a theory paper and a text book \cite{ref3,scullybk}.\\

\noindent
 The signal photons are sent to the detector $D_0$. The idler photons are sent to the beamsplitter setup. The path lengths in the experiment are arranged so that the signal photons reach detector $D_0$ before the idler photons  reach their final destination. So if the signal photon is detected at position $x$ on the screen, then our offer wave becomes \cite{ruth}
\be
\psi = \frac{\alpha}{\sqrt{2}} \left( \la x | A_s \ra | A_i \ra  +  \la x | B_s \ra  | B_i \ra   \right) \;\; .
\ee
A simple fourier transform of a slit with a constant electric field will give the single slit diffraction amplitude $\alpha$ in the form
\be
\alpha = \mbox{sinc} ( k_x a/2 )
\ee
where $a$ is the slit width and $k_x = k \sin \theta$  and the angle $\theta$ is the angular displacement from the center of the slits to the position $x$ on the screen. For the paraxial ray approximation this would be
\be
k_x = k \sin \theta = \frac{k x}{f} = \frac{\pi x }{\lambda f}  \label{kay}
\ee
where $f$ is the focal length of the lens which is taken to be roughly the slit screen distance and $\lambda$ is the wavelength of the signal photons and we have used $k = 2 \pi/\lambda$. Hence
\be
\alpha = \mbox{sinc} \left( \frac{ k x a}{2 f} \right) = \mbox{sinc} \left( \frac{ \pi x a}{ \lambda f} \right)  \label{al}
\ee

We will now assume that
\bea
\la x | A_s \ra &=& e^{ i k_x d_A} \nn \\
\la x | B_s \ra &=& e^{i k_x d_B }
\eea
where $d_A$ and $d_B$ are the distances from the crystal regions $A$ and $B$ to the screen at position $x$. Also we assume that the slit separation can be given by $d = d_A - d_B $. The offer wave can now be written as,
\be
\psi_{ow} = \frac{\alpha}{\sqrt{2}} \left(  e^{ i k_x d_A } | A_i \ra + e^{ i k_x d_B } | B_i \ra \right) \; .
\ee
Note that we have now dealt with the signal photons and only have to concern ourselves with the idler photon detection.  At this point we can continue with the Cramer interpretation or take the wave function Eq. (10) as a standard wave function and use spontaneous emission photon wave packets and expand them in terms or retarded and advanced waves to clearly see the overlap of the two and how the advanced waves retrace the retarded wave in time. This is carried out, for Case 1. below, in Appendix B. to show the technique.
Three cases follow:\\

\noindent
{\bf Case 1:}\\

\noindent
Assume the idler photon will be detected at detector $D_1$. The offer wave produced by passing photons through the beamsplitters will be
\be
\psi_{ow} = \frac{\alpha}{\sqrt{2}} \left(  e^{ i k_x d_A }  t r | A_i \ra + e^{ i k_x d_B } t^2 | B_i \ra \right)
\ee
the $A_i$ idler photon is transmitted through BSA and reflected from BS to reach $D_1$. The $B_i$ idler photon  is transmitted through BSB and transmitted through BS to reach $D_1$. See Fig. 4 for details of the paths. We have assumed that the extra path length
in traveling  through the beamsplitters is the same for both photons $A_i$ and $B_i$, otherwise we would need additional phase factors to account for the path length difference. The counter wave produced by detector $D_1$
will be the complex conjugate wave traveling backward in time towards the slits,

\be
\psi_{cw}^* = \frac{\alpha^*}{\sqrt{2}} \left(  e^{ -i k_x d_A }  t^* r^* \la A_i | + e^{ -i k_x d_B } t^{*2 } \la  B_i | \right) \; .
\ee
The probability that this transaction will occur then becomes,
\be
\psi_{cw}^* \psi_{ow} = \frac{1}{2} |\alpha |^2 \left[  |r|^2 |t|^2 \la A_i | A_i \ra + |t|^4 \la B_i | B_i \ra +
|t|^2  \left( r^* t \la A_i | B_i \ra e^{ -i k_x d} +  r t \la B_i | A_i \ra e^{i k_x d} \right) \right]
\ee
Let the amplitudes $\la A_i | A_i \ra = \la B_i | B_i \ra= 1 $, $\la A_i | B_i \ra = \eta_1^{1/2} \exp( -i \phi )$ and complex conjugate $\la B_i | A_i \ra = \eta_1^{1/2} \exp( i \phi )$ , where $\eta_1$ represents the detector efficiency of $D_1$ which is most likely less than unity.
The detector efficiency has been incorporated into the probability amplitude for convenience only.  Then we may write,
\be
\psi_{cw}^* \psi_{ow} = \frac{1}{2} |\alpha |^2  |t|^2 \left[  (|r|^2  + |t|^2 ) + \eta_1
  \left( r^* t  e^{ -i (k_x d + \phi) } +  r t  e^{i (k_x d + \phi ) } \right) \right]  \; .
\ee
Using our earlier results Eq(\ref{amps}) for the amplitudes $r$ and $t$ of the lossless beamsplitters and
\be
e^{\pm i \pi/2 } = \cos \pi/2 \pm i \sin \pi/2 = \pm i
\ee
we get,
\bea
\psi_{cw}^* \psi_{ow} &=&  \frac{1}{4} |\alpha |^2  \left[ 1 + \eta_1 \cos ( k_x d + \phi + \pi/2 ) \right] \nn \\
&=&  \frac{1}{4} |\alpha |^2  \left[ 1 + \eta_1 \cos \left( \frac{\pi d}{\lambda f} + \phi + \pi/2 \right) \right]  \label{new1}
\eea
It is more general to leave the result in this form. However the Kim paper \cite{shih} goes on to simplify further, uses $\eta_1 =1$ for perfect detection and writes,
\be
\psi_{cw}^* \psi_{ow} = \frac{1}{2} |\alpha |^2  \cos^2 \left[ \frac{k_x d}{2} + \frac{\phi}{2} + \frac{\pi}{4} \right]
\ee
where $\alpha$ is given by Eq.(\ref{al}) and $k_x$ is given by Eq.(\ref{kay}). In the last step we used the double angle formula for $\cos 2 \beta = 2 \cos^2 \beta - 1$. This is the coincidence result between detector $D_1$ together with detector $D_0$ and shows interference.\\

\noindent
Using our result Eq.(\ref{new1}) it is easy to generalize to a small spread of wavelengths (bandwidth=$\Delta \lambda$) by using a computer code to plot the equation and summing the interference patterns for $\lambda$, $\lambda \pm \Delta \lambda$,
$ \lambda \pm \Delta \lambda /2 $ and $ \lambda \pm \Delta \lambda /4 $. This will give a quite accurate interference pattern which will match the experimental data very well. If you also include the detector efficiency $\eta_1$ then you can match the experimental fringe visibility almost exactly. This is easy to do with a symbolic manipulation code like {\em Mathematica}, which also plots the results for you.
\\

\noindent
{\bf Case 2:}\\

\noindent
When the idler photons are detected at $D_2$ the offer wave becomes,
\be
\psi_{ow} = \frac{\alpha}{\sqrt{2}} \left(  e^{ i k_x d_A }  t^2 | A_i \ra + e^{ i k_x d_B } tr | B_i \ra \right)
\ee
Note that the $A_i$ photon is transmitted  by both BSA and BS, and the $B_i$ photon is transmitted by BSB but reflected by BS to reach $D_2$. See Fig. 4 for details.
The detector produces a counter wave which is the complex conjugate of the offer wave above,
\be
\psi_{cw}^* = \frac{\alpha^*}{\sqrt{2}} \left(  e^{ -i k_x d_A }  t^{*2} \la A_i | + e^{ -i k_x d_B } t^* r^* \la B_i |\right)
\ee
Using the same manipulations as before, leaving the detector efficiency as unity, the joint probability detection of coincidence counts between $D_0$ and $D_2$ becomes,

\bea
\psi_{cw}^* \psi_{ow} &=& \frac{|\alpha |^2 }{2}  |t|^2 \left[  |t|^2 \la A_i | A_i \ra + |r|^2 \la B_i | B_i \ra +
\left( t^* r \la A_i | B_i \ra e^{-ik_x d } + r^* t \la B_i | A_i \ra e^{i k_x d } \right) \right]  \nn \\
&=& \frac{|\alpha |^2 }{4} \left[ 1  + \frac{i}{2} e^{-i( k_x d + \phi)} - \frac{i}{2} e^{i( k_x d + \phi)} \right] \nn \\
&=& \frac{|\alpha |^2 }{4} \left[ 1 + \cos \left(  k_x d + \phi - \frac{\pi}{2}   \right)   \right] \nn \\
&=& \frac{|\alpha |^2 }{2} \cos^2 \left( \frac{k_x d}{2} + \frac{\phi}{2} - \frac{\pi}{4}   \right)
\eea
which also shows interference. The factor $\alpha$ is given by Eq.(\ref{al}). Note that this interference is $\pi$ out of phase with the interference pattern obtained from the coincidence count between $D_0$ and $D_1$. This is easier to see in the cosine result rather than the $\cos^2$ result. That means if the interference with $D_1$ is co-sinusoidal then this interference would be sinusoidal. This is exactly what was observed in the experiment \cite{shih}. \\

\noindent
{\bf Case 3:}\\

\noindent
If the idler photon is detected at either $D_3$ or $D_4$ then the corresponding offer waves would be,
\be
\psi_{ow} = \frac{ \alpha r }{\sqrt{2}}  \left( e^{i k_x d_A } | A_i \ra  + e^{i k_x d_B } | B_i \ra \right)
\ee
and the counter waves would be
\bea
\psi_{cw3}^* &=& \frac{ \alpha^* r^*}{\sqrt{2}} \la A_i | e^ {-i k_x d_A } \;\;\;  \mbox{  for detector $D_3$}  \nn \\
\psi_{cw4}^* &=& \frac{ \alpha^* r^*}{\sqrt{2}} \la B_i | e^ {-i k_x d_B } \;\;\; \mbox{   for detector $D_4$}
\eea
The probability of a coincidence count between $D_0$ and $D_3$ becomes,
\be
\psi_{cw3}^* \psi_{ow} = \frac{|\alpha|^2 |r|^2 }{2} \la A_i | A_i \ra = \frac{ |\alpha|^2}{4}
\ee
which shows no interference only a single slit diffraction pattern.
The probability of a coincidence count between $D_0$ and $D_4$ becomes,
\be
\psi_{cw4}^* \psi_{ow} = \frac{|\alpha|^2 |r|^2 }{2} \la B_i | B_i \ra = \frac{ |\alpha|^2}{4}
\ee
which likewise shows no interference. Again, the single slit diffraction amplitude $\alpha$ is given by Eq ( \ref{al}). This
also agrees with the experimental results of Kim et al. \cite{shih}.

\subsection*{Discussion}

\noindent
The transactional interpretation is related to the direct particle interaction theory of Wheeler -- Feynman and Hoyle -- Narlikar and involves advanced waves as well as the usual retarded waves. The advanced waves are natural solutions to the relativistic wave equation and are required to conserve momentum in direct particle interactions. This paper has briefly considered the pros and cons of direct particle interactions verse conventional field theory methods. In terms of vacuum energy density the direct particle approach tells us there is no vacuum field and thus its energy is identically zero, close in fact to the observed value. Quantum field theory tells us that the vacuum energy density is huge and gives a value 120 times too large. Direct particle or source theory does away with self interaction and subtracting infinities is only needed for charge renormalization. Charge renormalization follows in the same manner as in the field theory case when you introduce a size cutoff (no point particles) of the Schwarzschild radius of the particle. There is also a size limit to the universe to prevent a divergent advanced wave integral due to the Rindler horizon for an accelerating expansion of the universe \cite{me}. Advanced waves have never been detected in practice and this lack of experimental evidence is enough for some to rule them out altogether. \\

\noindent
It only takes one experimental observation to refute a theory. John Cramer and Nick Herbert \cite{nick2} considered several experimental possibilities of nonlocal quantum signaling (retrocausal signals) involving path entangled systems and in all cases found that the complementarity between two-photon interference and one-photon interference blocks any potential nonlocal signal \cite{jaeger}.
 The traditional way of thinking about an instantaneous wave function collapse, at a certain time at a certain place, which is clearly in conflict with relativity, is superseded in the transactional picture. The wave function collapse is among the most confusing aspects of quantum mechanics (as a component of the measurement problem) and is simply resolved using the TI method of Cramer, or PTI of Kastner. Indeed the Copenhagen approach actually evades the entire issue by taking the wave function and its collapse as epistemic--a measure of knowledge rather than a physical entity. This approach is observer-dependent; it is subject to the 'Heisenberg Cut' in which there is no physically grounded and non-arbitrary account of what constitutes an 'observer'. In the transactional approach, there is no observer-dependence: it is absorbers that provide the missing ingredient that defines when a measurement and attendant collapse occurs. \\

\noindent
Advanced waves are natural solutions to relativistic wave equations. In order to use this theory for the nonrelativistic case it is necessary to think of two Schr\"odinger equations:  one Schr\"odinger equation for the wave function $\psi$ and one for its complex conjugate $\psi^*$, which becomes the advanced wave. This makes sense if we think of the Schr\"odinger equation as a square root version of the relativistic Klein Gordon equation.\\

\noindent
Furthermore, work by Hogarth \cite{hogarth} and Hoyle and Narlikar (HN) \cite{HN64, HN, HNbook} has paved the way to a
new version of direct particle interaction gravitational theory, which is fully Machian, incorporates advanced waves and has Einstein's theory as a special case. The HN theory may be quantized as in their book \cite{HNbook} using the path integral technique pioneered by Feynman \cite{space48}.\\

\noindent
It is interesting to note that the mass field $m(x)$ in HN theory looks similar to the source field $S(x)$ introduced by Schwinger \cite{schwinger1}. Wheeler  never gave up on the absorber theory, which is a direct particle interaction (action--at--a--distance) theory. It simply wasn't popular at the time and dropped off the radar. Gerard t'Hooft found a way to renormalize Yang Mills field theories in a way similar to QED and most physicists took that path. We believe the works of Cramer, Wheeler--Feynman, Hoyle--Narlikar, and Schwinger's source theory, are all direct particle interactions. How source theory is related to the Feynman path integrals is explained by Schweber \cite{schweber}. It should be noted that Schwinger was able to derive the Casimir force using the source field method in which there are no nontrivial vacuum fields \cite{peter,schwinger2}. The action at a distance theories are well worth study and may lead to a consistent picture of quantum gravity. Radiation reaction can be dealt with using the half retarded half advanced absorber picture. Many QED results thought to be vacuum fluctuation related can in fact be derived by considering source fields instead, including the Lamb shift and particle self energy \cite{peter}. 

\subsection*{Conclusions}

The main aim of this paper is to draw attention to the fact that the transactional interpretation of quantum mechanics by John Cramer is perfectly viable and legitimate, and should be given due consideration by the physics community, which has not been the case thus far.   
The TI by Cramer \cite{cramer}, gives a simple and intuitive picture for wave function collapse distributed over the entire path of the interacting system (in Kastner's approach, the collapse is what establishes that path). In the case of the Kim experiment \cite{shih}, the wave function would collapse along the entire path between the slits (or the regions $A$ and $B$ of the down converting crystal) and the detectors and it would happen in a way distributed over time, not in an instant. The TI picture rules out the possibility of any backward in time signals using quantum delayed choice experiments. In fact it makes clear the idea is nonsense since the advanced counter wave from the detector must travel the entire distance back to the slit in order for the photon (from the slit) to make the trip in the first place. The choice is really no longer delayed since the photon {\em knows} where it will end up because of the advanced wave coming backwards in time to confirm the interaction or {\em handshake}, as Cramer puts it. The alternative way of avoiding wave function collapse is to use the correlation functions as in Appendix A. The calculations are far more long winded, than the fairly quick and easy calculation in the main paper, and in the opinion of the author the correlation function method masks what is really going on and thus leaves room for misinterpretation.\\

\subsection*{Appendix A.}

\noindent
Here we derive the Shih experimental result via the usual quantum optics correlation function approach and show the steps omitted in the experimental paper, \cite{shih}.
You could approximate the parametric down conversion photons with spontaneous emission photons and use the results in the Scully Druhl paper \cite{druhl}. This would give a sensible answer, but we have given the parametric downconversion theory in detail in what follows.
The quantum mechanical interaction picture Hamiltonian for the non-degenerate parametric downconversion in the rotating wave approximation \cite{scullybk} is
\be
V_{\mbox{\tiny int}} = \hbar \kappa ( \ad_s \ad_i a_p + a_s a_i \ad_p )
\ee

where $\ad_s$, $\ad_i$ and $\ad_p$ are the creation operators for the signal, idler and pump beams respectively and $a_s$, $a_i$ and $a_p$ are the corresponding annihilation operators.  The  coupling constant $\kappa$  depends on the second order susceptibility tensor which mediates the interaction, \cite{scullybk}.
In the non degenerate operation we find a two mode squeezed state output. In degenerate operation, where the signal and idler frequencies are the same and each half the pump frequency, you would get a single mode squeezed state. In the parametric approximation, the pump beam is treated classically as a coherent state and pump depletion can be neglected. If we allow $\alpha_p$ and $\theta$ to be the real amplitude and phase of the pump then the interaction Hamiltonian becomes,
\be
V_{\mbox{\tiny int}} = \hbar \kappa  \alpha_p ( \ad_s \ad_i  e^{-i \theta} + a_s a_i e^{i \theta} )
\ee
The equation of motion for the signal annihilation operator, taking the expectation over the signal vacuum becomes ;
\bea
\dot{a}_s &=& \frac{i}{\hbar} \la 0 | [ V_{\mbox{\tiny int}}, a_s ] |0 \ra_s  \nn \\
 &=& -i \Omega_p \ad_i e^{-i \theta} 
 \eea
  where $\Omega_p =  \kappa \alpha_p$. The signal creation operator equation of motion becomes $\dot{\ad}_s  = i \Omega_p a_i e^{i \theta}$.
  
  \noindent
 Similarly for the idler operators we use the idler vacuum to find; 
 \bea
 \dot{a}_i &=& -i \Omega_p \ad_s e^{-i \theta} \nn \\
  \dot{\ad}_i &=& i \Omega_p a_s e^{i \theta} 
  \eea
  By differentiating the above equations with respect to time and substitution we can find,
 \bea
 a_s (t) &=& A_s \cosh ( \Omega_p t ) + B_s \sinh ( \Omega_p t ) \nn \\
 \ad_s (t) &=& A^{\dagger}_a \cosh ( \Omega_p t ) + B^{\dagger} _s  \sinh ( \Omega_p t ) 
 \eea
 from which you can set $t=0$, and find solutions for the initial conditions. By substituting back the original equations, you can easily find the $A_s$, $B_s$ coefficients in terms of initial conditions for the creation and annihilation operators as follows,
 \bea
 a_s (t) &=& a_s (0) \cosh ( \Omega_p t ) - i e^{-i \theta} \ad_i (0) \sinh( \Omega_p t ) \nn \\
 \ad_s (t) &=& \ad_s (0) \cosh ( \Omega_p t) +i e^{i \theta} a_i (0) \sinh (\Omega_p t )  \; .
 \eea
 Similarly for the idler operators,
 \bea
 a_i (t) &=& a_i (0) \cosh ( \Omega_p t) - i e^{-i\theta} \ad_s (0) \sinh ( \Omega_p t ) \nn \\
 \ad_i (t) &=& \ad_i (0) \cosh (\Omega_p t ) + i e^{i \theta} a_s (0) \sinh ( \Omega_p t) \; .
 \eea
 
 \noindent
 For $\theta = \pi/2$ these look like non degenerate squeezed state transformations, \cite{scullybk}.
 For simplicity we are using type I parametric down conversion and degenerate frequencies. The frequency of the pump is the sum of the signal and idler frequencies. The signal and idler frequencies are taken to be the same. $\omega_p = \omega_s + \omega_i$, where $\omega_s = \omega_i$. 
In type I parametric downconversion the polarization of the signal and idler are the same.  In the experiment \cite{shih}, the signal photons interfere and the idler photons interfere separately so it makes no difference that they are from type II parametric down conversion and thus in perpendicular polarization states. We shall also use the same simplifying assumptions as in the previous transactional interpretation method. We assume that the separation of the region A and B from the detector $D_0$ are very similar the only difference in path length being the region separation. We further assume that
 the idler distances from region A or B to the same detector $D_1$ -- $D_4$ are the same. This brings about a great simplification in that the integrations are over 2 times and not 4.
 The extra work involved in allowing the signal photons to have two distinct path lengths and the two idler photons to also have two distinct path lengths, to the same detector, does not add to the physics and only complicates the integrations unnecessarily. This is easy to set up but gets messy, very quickly, in practice. \\
 
 \noindent
 {\bf Joint Detection $D_0$ and $D_1$ detectors}\\
 
 \noindent
 For the probability of joint detection $R_{0,1}$ from detectors $(D_0 ,D_1)$ we set up the following integration \cite{shih},
 
 \be
 R_{01} \propto \frac{1}{T} \int_0^T  \int_0^T dt_0 dt_1 \la : E^{(-)}_s ( t_0 ) E^{(+)}_s (t_0 ) E^{(-)}_i ( t_i ) E^{(+)}_i  (t_i )  : \ra
 \ee
 where $ \la : \;\; : \ra $ denotes normal ordering where all creation operators are to the left of all the annihilation operators. The $i$ will take values of 1-4 depending on the idler detector $D_1 $-- $D_4$. Here $t_0$ is the time for the signal photons to go from the crystal to the detector $D_0$ and $t_1$ is the time for the idler photons to get from the crystal to detector $D_1$. We take the signal path length to be $d_A$ or $d_B$ for the two regions and the idler path length to be $x_A$ and $x_B$ from the crystal to detector one. From the experiment $t_0 < t_1$ by about 8ns.
Shih et al \cite{shih} tell us that the above integral is approximately the same as the integral of  $ |\la E^{(+)} (t_0 ) E^{(+)} ( t_1 )  \ra |^2 $.  
The positive frequency part of the electric signal is
 $ E^{(+)}_s (t) = E_0 a_s (t) e^{i \omega_s t }$ the negative part is $ E^{(-)}_s (t) = E_0 \ad_s (t) e^{-i \omega_s t }$, where $E_0$ is some constant. The interference results are usually normalized so we set $E_0 =1$ in what follows.
 We drop all the $\omega$ terms , $\omega_p = \omega_s + \omega_i$ since they will all cancel out, and we take $\omega_s = \omega_i$ for simplicity.
Actually if you expand the 4th order correlation function you get 3 such terms as follows, see Collett and Loudon \cite{collett};

\bea
\la : E^{(-)}_s ( t_0 ) E^{(+)}_s (t_0 ) E^{(-)}_i ( t_1 ) E^{(+)}_i (t_1 )  : \ra  & =&  \la E^{(-)}_s (t - t_0) E^{(+)}_i (t' - t_1)\ra \la E^{(-)}_i (t' -t_1 ) E^{(+)}_s ( t-t_0 )\ra \nn \\
 & +& \la  E^{(-)}_s ( t -t_0 )  E^{(-)}_i  ( t' -t_1 ) \ra \la  E^{(+)}_s (t - t_0 ) E^{(+)}_i  ( t' - t_1 ) \ra \nn \\
 & +& \la E^{(-)}_s (t-t_0  ) E^{(+)}_s  ( t - t_0 ) \ra \la E^{(-)}_i  ( t' - t_1 ) E^{(+)}_i  ( t' -t_1) \ra \nn \\
 \eea

 It turns out only the first term cancels but the other two terms are non zero. Collett and Loudon  \cite{collett} outline a more advanced time integration procedure. We are approximating with two times only assuming the distances for both signal photons are almost the same and the idler photons have equal path lengths to the same detector.
 The signal and idler electric fields for detection at $D_0$ and $D_1$ can be written as;
 \bea
 E^{(+)}_s(t) &=& \sqrt{ \frac{\alpha}{2} } \left( a_s e^{i k_x d_A }  \cosh ( \Omega_p t_0 ) - i \ad_i t^{\star}r^{\star} e^{-i k x_A } e^{-i \theta} \sinh ( \Omega_p t_1 )  \right) \nn \\
  & & + \sqrt{ \frac{\alpha}{2} }  \left(  a_s e^{i k_x d_B }  \cosh ( \Omega_p t_0 ) - i \ad_i t^{\star 2} e^{-i k x_B } e^{-i \theta} \sinh ( \Omega_p t_1 )  \right) \nn \\
  E^{(+)}_{i=1} (t) &=& \sqrt{ \frac{\alpha}{2} } \left( a_1  r t e^{i k x_A }  \cosh ( \Omega_p t_1 ) - i \ad_s e^{-i k_x d_A } e^{-i \theta} \sinh ( \Omega_p t_0 )  \right) \nn \\
  & & + \sqrt{ \frac{\alpha}{2} }  \left(  a_1  t^2 e^{i k x_B }  \cosh ( \Omega_p t_1 ) - i \ad_s  e^{-i k_x d_B } e^{-i \theta} \sinh ( \Omega_p t_0 )  \right)
  \eea
 where the first line of each electric field equation is from region $A$ of the crystal, and the second line comes from region $B$. The $\alpha$ term is the sinc function or the square root of the single slit diffraction pattern as  defined in the TI section. See Eq.s (6-9) in this paper. The expectation values are evaluated in a vacuum.
 After some tedious algebra it can be shown that the first term in Eq. (33) gives zero. The only non-zero terms have combinations of $\la 0| a_s  \ad_s |0 \ra$ , 
 $\la 0| a_i \ad_i |0 \ra$  in them. The second order correlation functions in the second term are; 
 \bea
  \la E^{(+)}_s E^{(+)}_1 \ra &=& -i e^{-i \theta} \la a_s \ad_s\ra \frac{\alpha}{2} 2 \sinh(\Omega_p t_0 ) \cosh( \Omega_p t_0 ) ( 1 + \cos[ k_x d] )   \nn \\
  \la E^{(-)}_s E^{(-)}_1 \ra &=& i e^{i \theta} \la a_1 \ad_1 \ra \frac{\alpha}{2} |t|^2 \sinh(\Omega_p t_1 ) \cosh( \Omega_p t_1 ) \left[ |r|^2 + |t|^2  +
  r^{\star}t e^{-ik(x_A - x_B )}  + rt^{\star} e^{ik(x_A - x_B )}  \right] \nn \\
  \eea
where we have used the lossless beamsplitter result that $ rt^{\star} + r^{\star} t =0$ and $|r|^2 + |t|^2 =1$ and $d = d_A - d_B $ is the slit separation 
(distance between regions $A$ and $B$ or the crystal). 
 It is also assumed that $x_A = x_B$ so the idler photons travel the same distance to the same detector $D_1$. 
 The second term in the expansion with $i = 1$ for $D_1$ becomes;
 \bea
 \la  E^{(+)}_1 E^{(+)}_s  \ra \la  E^{(-)}_1 E^{(-)}_s  \ra = \frac{\alpha^2}{4} &&  \!\!\!\!\! \!\!\!\! \cosh( \Omega_p t_0 ) \sinh (\Omega_p t_0 )\cosh( \Omega_p t_1) 
 \sinh (\Omega_p t_1 ) \nn \\ 
 && \times 2 |t|^2( 1 + \cos[ k_x d] ) \; .
 \label{term2}
 \eea
Similarly,
\bea
  \la E^{(-)}_s E^{(+)}_s \ra &=&   \la a_1 \ad_1\ra \frac{\alpha}{2}  \sinh^2 (\Omega_p t_1 ) |t|^2  \left[   (|r|^2 + |t|^2)  +  r^{\star}t + rt^{\star}  \right] \nn \\
  \la E^{(-)}_1 E^{(+)}_1 \ra &=&   \la a_s \ad_s \ra \frac{\alpha}{2}  2 \sinh^2(\Omega_p t_0 ) ( 1 + \cos[ k_x d] )    
 \eea 

\noindent   
 The third term in the expansion Eq. (33)  becomes;
  \bea
 \la  E^{(-)}_s E^{(+)}_s \ra  \la E^{(-)}_1 E^{(+)}_1 \ra &=& \frac{\alpha^2}{2} \sinh^2( \Omega_p t_0 ) \sinh^2 ( \Omega_p t_1 ) |t|^2 \nn \\
 & & \times ( 1 + \cos[ k_x d] ) 
 \label{term3}  
  \eea
Hence, adding terms 2 , Eq. (\ref{term2}) and term 3, Eq. ( \ref{term3}) we find the probability $R_{01}$ to be,
 \bea
 \la : E^{(-)}_s ( t_0 ) E^{(+)}_s (t_0 ) E^{(-)}_1 ( t_1 ) E^{(+)}_1 (t_1 )  : \ra   \propto &&   \!\!\!\!\!\! \frac{ \alpha^2}{2}  |t|^2 \sinh ( \Omega_p t_0 ) \sinh ( \Omega_p t_1 ) \nn \\
 && \times  \cosh ( \Omega_p [ t_0 + t_1])  ( 1 + \cos [ k_x d ] ) \; .
 \eea
  
 \noindent  
 The $ \frac{1}{T} \int_0^T \cosh^2 ( \Omega_p t_0 ) dt_0 $ and $ \frac{1}{T} \int_0^T \cosh^2 ( \Omega_p t_1 ) dt_1$  integrals, can be performed  
 and lead to constants so long as $ \Omega_p T > 0$. Clearly the $\cos[ k_x d]$ term leads to interference of the signal photons.\\

\vspace{0.2 in}

\noindent
 {\bf Joint Detection $D_0$ and $D_2$ detectors}\\

 \noindent
 The joint probability $R_{0,2}$, detection of $(D_0, D_2)$ leads to similar interference terms. The starting electric fields for detector 2 become;
\bea
 E^{(+)}_s(t) &=& \sqrt{ \frac{\alpha}{2} } \left( a_s e^{i k_x d_A }  \cosh ( \Omega_p t_0 ) - i \ad_2 t^{\star 2} e^{-i k x_A } e^{-i \theta} \sinh ( \Omega_p t_2 )  \right) \nn \\
  & & + \sqrt{ \frac{\alpha}{2} }  \left(  a_s e^{i k_x d_B }  \cosh ( \Omega_p t_0 ) - i \ad_2 t^{\star } r^{\star} e^{-i k x_B } e^{-i \theta} \sinh ( \Omega_p t_2 )  \right) \nn \\
  E^{(+)}_2 (t) &=& \sqrt{ \frac{\alpha}{2} } \left( a_2  t^2 e^{i k x_A }  \cosh ( \Omega_p t_2 ) - i \ad_s e^{-i k_x d_A } e^{-i \theta} \sinh ( \Omega_p t_0 )  \right) \nn \\
  & & + \sqrt{ \frac{\alpha}{2} }  \left(  a_2  t r  e^{i k x_B }  \cosh ( \Omega_p t_2 ) - i \ad_s  e^{-i k_x d_B } e^{-i \theta} \sinh ( \Omega_p t_0 )  \right)
  \eea
  \noindent
Since we have chosen to calculate type I, there will be no polarization change and we expect a similar result to that of $R_{0,1}$ above with the only difference that
$t_1 \rightarrow t_2$. We have not worried about any subtle phase changes on reflection here.\\

\vspace{0.2 in}

\noindent
 {\bf Joint Detection $D_0$ and $D_3$ detectors}\\

\noindent
The joint probability $R_{0,3}$ ,  detection of  $(D_0 , D_3)$ signal and idler photons can be calculated using a similar technique but the starting electric fields would be, using $i=3$;
\bea
 E^{(+)}_s(t) &=& \sqrt{ \frac{\alpha}{2} } \left( a_s e^{i k_x d_A }  \cosh ( \Omega_p t_0 ) - i \ad_3 r^{\star} e^{-i k x_A } e^{-i \theta} \sinh ( \Omega_p t_3 )  \right) \nn \\
 E^{(+)}_3 (t) &=& \sqrt{ \frac{\alpha}{2} } \left( a_3  r  e^{i k x_A }  \cosh ( \Omega_p t_3 ) - i \ad_s e^{-i k_x d_A } e^{-i \theta} \sinh ( \Omega_p t_0 )  \; .\right) \nn \\
  \eea 
 In this case only idler photons from region $A$ can reach detector 3. This implies that the signal photons also came from region $A$ and no interference results. 
 The new term 2 becomes;
 \bea
  \la E^{(+)}_s E^{(+)}_3 \ra &=& \la  a_s \ad_s  \ra \frac{\alpha}{2} ( -i e^{-i \theta} ) \cosh( \Omega_p t_0 ) \sinh ( \Omega_p t_0 ) \nn \\
  \la E^{(-)}_s E^{(-)}_3 \ra &=& \la   a_3 \ad_3  \ra \frac{\alpha}{2}  ( i e^{i \theta} ) \cosh (\Omega_p t_3 ) \sinh( \Omega_p t_3 ) |r|^2 \nn \\
  \la E^{(+)}_s E^{(+)}_3 \ra  \la E^{(-)}_s E^{(-)}_3 \ra &=& \frac{\alpha^2}{4} \cosh (\Omega_p t_0 ) \sinh (\Omega_p t_0 ) \cosh (\Omega_p t_3 ) \sinh ( \Omega_p t_3) |r|^2 \; .
  \eea
  The new term 3 becomes;
  \bea
  \la E^{(-)}_s E^{(+)}_s \ra &=& \la  a_3 \ad_3  \ra \frac{\alpha}{2} |r|^2 \sinh^2 ( \Omega_p t_3 ) \nn \\
  \la E^{(-)}_3 E^{(+)}_3 \ra &=& \la   a_s \ad_s  \ra \frac{\alpha}{2}  \sinh^2 ( \Omega_p t_0 )  \nn \\
  \la E^{(-)}_s E^{(+)}_s \ra  \la E^{(-)}_3 E^{(+)}_3 \ra &=& \frac{\alpha^2}{4} |r|^2  \sinh^2 (\Omega_p t_0 ) \sinh^2 ( \Omega_p t_3)  \; .
  \eea 
  The point probability  $R_{03}$ becomes ;
  \be
  \la : E^{(-)}_s ( t_0 ) E^{(+)}_s (t_0 ) E^{(-)}_3 ( t_3 ) E^{(+)}_3 (t_3 )  : \ra   \propto  \frac{\alpha^2}{4} |r|^2 \sinh^2 ( \Omega_p t_0 )\sinh^2 ( \Omega_p t_3 )  
  \cosh( \Omega_p [ t_0 +t_3 ])
  \ee
  Clearly no interference present.\\
 
 \vspace{0.2 in} 
  
\noindent
 {\bf Joint Detection $D_0$ and $D_4$ detectors}\\

 \noindent
 The joint probability $R_{0,4}$, detection of $(D_0 , D_4)$ signal and idler photons $i=4$, can be calculated using the electric fields below;
\bea
 E^{(+)}_s(t) &=& \sqrt{ \frac{\alpha}{2} } \left( a_s e^{i k_x d_B }  \cosh ( \Omega_p t_0 ) - i \ad_4 r^{\star} e^{-i k x_B } e^{-i \theta} \sinh ( \Omega_p t_4 )  \right) \nn \\
 E^{(+)}_4 (t) &=& \sqrt{ \frac{\alpha}{2} } \left( a_4  r  e^{i k x_B}  \cosh ( \Omega_p t_4 ) - i \ad_s e^{-i k_x d_B } e^{-i \theta} \sinh ( \Omega_p t_0 )  \; .\right) \nn \\
  \eea 
  Only idler photons from region $B$ can reach detector 4. This implies the signal photons came from region $B$ also, and so no interference. The joint probability $R_{0,4}$ is very similar to the previous result for $R_{03}$ with $t_3 \rightarrow t_4$.\\
  

\subsection*{Appendix B.}

\noindent
Here we derive the results for Case 1, treated in the main paper, but using a symmetric wavefunction with both retarded and advanced waves. Using the notation from the book by Zubairy and Scully \cite{scullybk} we find that a spontaneously emitted photon (idler photon in our case) can be represented by a wave function of the type,
\be
\la 0| E^{+} | \phi_i \ra = -i  \frac{ \wp_{ab} \sin \eta }{ 8 \epsilon_0 \pi^2 \Delta r} \frac{\w^2}{c^2} \int_{-\infty}^{\infty}  d\nu_k \left[
\frac{ e^{-i \nu_k t + i \nu_k \Delta r/c} }{ \nu_k -\w +i \gamma/2 } - \frac{ e^{-i \nu_k t - i \nu_k \Delta r/c} }{ \nu_k -\w +i \gamma/2 } \right]
\ee
Using the contour integration in \cite{scullybk} the upper hemisphere anti clockwise gives zero since there is no pole, the lower hemisphere clockwise gives a simple residue at $ \nu_k =\w - i\gamma/2$. This gives the result,
\bea
\la 0| E^{+} | \phi_i \ra &=& \varepsilon_0 \left[ e^{ (-i \w - \gamma/2 ) ( t - \Delta r/c) } \theta ( t - \Delta r /c ) -e^{ (-i \w - \gamma/2 ) ( t + \Delta r/c) } \theta ( t + \Delta r /c ) \right] \nn \\
\varepsilon_0 &=& \left( \frac{ \w^2 \wp_{ab} \sin \eta }{ 4 \pi \epsilon_0 \Delta r c^2 } \right) \label{photon}
\eea
where the spontaneous decay is $\gamma $, the atomic transition dipole matrix element is $\wp_{ab}$ and $\eta$ is the angle between the dipole matrix element and the z--axis. The frequency $\w$ is the idler frequency. The 
$\theta(t \pm \Delta r/c)$ functions are determined from the direction around the contour integration taken to find a nonzero result. The negative sign is for retarded waves the positive sign is for the advanced waves going backward in time. The Eq. (\ref{photon}) is used for both idler photons for the Case 1. Starting from Eq. (10) in the main text the wave function 
for the idler photon to be detected by detector 1 becomes,
\bea
\psi_1 &=& \frac{\alpha \varepsilon_0}{\sqrt{2} } \left[ e^{(-i \w - \gamma/2)(t - L_{1A}/c)}e^{ik_x d_A } \theta( t - L_{1A}/c ) - 
e^{(-i \w - \gamma/2)(t + L_{1A}/c)}e^{ik_x d_A } \theta( t +L_{1A}/c )  \right.    \nn \\
& & \left. + e^{(-i \w - \gamma/2)(t - L_{1B}/c)}e^{ik_x d_B } \theta( t - L_{1B}/c ) - e^{(-i \w - \gamma/2)(t + L_{1B}/c)}e^{ik_x d_B } \theta( t + L_{1B}/c ) \right]  \nn \\
& &
\eea
where we have approximated by missing out the $r$ and $t$ reflection and transmission coefficients. These would lead to a numerical factor and possibly a phase shift which is not of importance at the moment. (Note -- this is to eliminate any confusion between the transmission coefficient and the time $t$.)
The lengths from region A,B of the crystal to detector 1 are $L_{1A} \; ,  L_{1B} $ respectively. For interference we want to find $\psi_1^{\star} \psi_1$. It is quite straightforward to multiple this out. For convenience we make the further simplifying assumptions;
\bea
L_{1A} & \approx & L_{1B} = L \nn \\
\frac{L_{1A}-L_{1B} }{c} & \approx & \delta t
\eea
It is assumed that the path lengths from the regions A,B of the crystal to detector 1 are almost the same and equal to length $L$, which could be a meter or more in length.
It is further assumed, that if there is a path difference from regions A,B of the crystal to the detector 1, it is very small so that the path difference divided by $c$ becomes
$\delta t \rightarrow 0$. The following result is then found for $\psi_1^{\star}\psi_1$,
\bea
\psi_1^{\star} \psi_1 &=&  | \alpha |^2 \varepsilon_0^2 \left\{    e^{-\gamma (t - L/c)} \theta^2(t-L/c) +  e^{-\gamma (t +L/c)} \theta^2(t+L/c)  \right. \nn \\
 &+&   \left. \cos [ \w \delta t + k_x d ] e^{ - \gamma( t - L/c )} \theta^2(t-L/c) + \cos [ \w \delta t - k_x d ] e^{ - \gamma( t + L/c)} \theta^2(t +L/c)  \right. \nn \\
 &+& \left.  \cos [ 2L\w /c + k_x d ]  \left[ e^{ -\gamma (t - \delta t/2 ) } + e^{- \gamma ( t + \delta t /2)} \right] \theta( t - L/c) \theta( t+ L/c)  \right. \nn \\
 &-& \left. 2 e^{-\gamma t} \cos (2 \w L/c)   \theta( t - L/c) \theta( t+ L/c)        \right\}
 \eea
 where $d = d_A - d_B$ as before. The result is symmetric in the retarded and advanced waves. The advanced waves are normally not detectable. The first line shows single slit diffraction terms. These theta squared terms were just in lengths for paths $L_{1A}$ or $L_{1B}$ alone and a factor of 2 has been removed. The interference is clear from the second line of the above equation. This results from a path interference between lengths $L_{1A}$ and $L_{1B}$. Both terms are either retarded or both advanced. The 3rd and 4th lines show an interference between the retarded and advanced waves. The 3rd line is actually a mixture of theta functions from paths $L_{1A}$ and $L_{1B}$, the 4th line was originally two terms, one from region $A$ and the other from region $B$. The full expression is rather long, so both arm lengths from crystal to detector 1 were taken to be approximately the same length $L$. The value of $2L\w/c$ can be very large of order $\sim 10^7$ for lengths $L$ of a meter, and frequency $\w = 3 \times 10^{15} $rad/s. Interference of the retarded and advanced waves takes place along the entire path length $L$. An advanced wave returns along the same path as the outgoing retarded wave, but the advanced wave travels in the reverse time direction from detector to slits and thus collapses the wave function along the entire path of the photon.  The last term would most likely not be visible due to the large argument of the cosine which would have a tendency to cause rapid oscillation and wash out the fringes as a result (for any variation in $\w$). This appears to confirm Cramer's hypothesis that the wave function collapse is not instantaneous, but is distributed in time along the flight path of the photon. 
 
   
\subsection*{Acknowledgements}

\noindent
HF thanks both J. G. Cramer and R. E. Kastner for reviewing the paper and for suggesting great improvements. Thanks also go to K. Wanser and P. W. Milonni for helpful comments.


\end{document}